\documentclass[conference]{IEEEtran}

\usepackage{cite}
\usepackage{amsmath,amssymb,amsfonts}
\usepackage{algorithmic}
\usepackage{graphicx}
\usepackage{hyperref}
\usepackage{listings}
\usepackage{textcomp}
\usepackage{xcolor}
\usepackage{pgfplots}
\usepackage{siunitx}

\lstdefinestyle{codestyle}{
  keywordstyle=\color{blue},
  stringstyle=\color{red},
  commentstyle=\color{lightgray},
  morecomment=[l][\color{codegreen}]{\#},
  basicstyle=\fontsize{6}{6}\selectfont\ttfamily\color{black},
  breakatwhitespace=false,         
  breaklines=true,                 
  captionpos=b,                    
  keepspaces=true,                     
  showspaces=false,                
  showstringspaces=false,
  showtabs=false,                  
  tabsize=1
}
\begin{document}
\title{A C++20 Interface for MPI 4.0}

\makeatletter
\newcommand{\linebreakand}{%
  \end{@IEEEauthorhalign}
  \hfill\mbox{}\par
  \mbox{}\hfill\begin{@IEEEauthorhalign}
}
\makeatother

\author{\IEEEauthorblockN{Ali Can Demiralp}
\IEEEauthorblockA{\textit{Visual Computing Institute} \\
\textit{RWTH Aachen University}\\
Aachen, Germany \\
demiralp@vis.rwth-aachen.de}
\and
\IEEEauthorblockN{Philipp Martin}
\IEEEauthorblockA{\textit{IT Center} \\
\textit{RWTH Aachen University}\\
Aachen, Germany \\
pm.martin@itc.rwth-aachen.de}
\and
\IEEEauthorblockN{Niko Sakic}
\IEEEauthorblockA{\textit{IT Center} \\
\textit{RWTH Aachen University}\\
Aachen, Germany \\
sakic@itc.rwth-aachen.de}
\linebreakand
\IEEEauthorblockN{Marcel Kr\"uger}
\IEEEauthorblockA{\textit{Visual Computing Institute} \\
\textit{RWTH Aachen University}\\
Aachen, Germany \\
krueger@vis.rwth-aachen.de}
\and
\IEEEauthorblockN{Tim Gerrits}
\IEEEauthorblockA{\textit{Visual Computing Institute} \\
\textit{RWTH Aachen University}\\
Aachen, Germany \\
gerrits@vis.rwth-aachen.de}
}

\maketitle

\begin{abstract}
We present a modern C++20 interface for MPI 4.0.
The interface utilizes recent language features to ease development of MPI applications.
An aggregate reflection system enables generation of MPI data types from user-defined classes automatically.
Immediate and persistent operations are mapped to futures, which can be chained to describe sequential asynchronous operations and task graphs in a concise way.
This work introduces the prominent features of the interface with examples.
We further measure its performance overhead with respect to the raw C interface.
\end{abstract}


\section{Introduction}
The message passing interface (MPI) is the standard programming model for distributed computing today, yet it lacks an official C++ interface since version 3.0.
Applications written in C++ have to rely on the C interface, which provides no encapsulation, requires manual memory and scope management, and prevents use of C++ idioms and features.
As detailed by R\"ufenacht et al. \cite{Ruefenacht2021}, many unofficial C++ interfaces such as Boost.MPI \cite{Gregor2005} and MPL \cite{Rabauke2015, Ghosh2021} exist.
However these often target earlier versions of MPI $(<4.0)$, cover a subset of their respective specification, and tend to limit their usage of language features, serving mostly as RAII wrappers.
Although the MPI forum is actively discussing the potential features of a new C++ interface \cite{MPI2020}, it is currently far from standardization. 

This work presents a modern C++20 interface for MPI 4.0, covering the complete specification.
The interface provides automatic lifetime management for each MPI object, meaningful defaults for each MPI function, compile-time generation of MPI data types from structures and classes, and the ability to express MPI requests as futures with continuations to describe sequential non-blocking communication.
We implement the majority of the features requested by the users in \cite{MPI2020}, envisioning what an official C++ interface could look like today.

\section{Implementation}
The interface is implemented as a header-only library depending on the C interface.
It consists of three major components; the core, IO and tool interfaces. 
The core component implements the chapters 1-13 of the standard \cite{MPI2021}, containing all point-to-point, collective and one-sided communication.
The IO component implements chapter 14, the MPI-IO interface, covering the functions with the prefix \textit{MPI\_File\_}.
The tool component implements chapter 15, the profiling and tool information interface of MPI, containing the functions with the prefix \textit{MPI\_T\_}.

The interface closely follows the C++ core guidelines \cite{CPP2015}, which encourage idiomatic use of the language and the standard library.
The classes have two types of constructors; managed and unmanaged.
Managed constructors instantiate a new MPI object and assume responsibility for its destruction.
Unmanaged constructors accept an existing MPI object and do not assume responsibility for its destruction by default.
Copy constructors are deleted unless MPI provides duplication functions (ending with \textit{\_dup}) for the object, whereas move constructors are available whenever possible.

All function pointers are converted to \textit{std::function}s, which enables user data to be passed through captures rather than void pointer arguments.
The library further contains scoped versions of each enumeration.
Functions expecting enumerations as arguments use these scoped versions, which prevent passing erroneous values and provide code completion support.
The arguments of functions that accept a variety of MPI objects are described with a \textit{std::variant}, providing constrained type erasure. 
Optional arguments and indeterminate return values, such as the result of an immediate probe, are described using \textit{std::optional}. 

The library provides default arguments for most MPI functions, according to the standard where applicable.
In several cases, the defaulted arguments are required to be moved at the end of the argument list.
Furthermore, functions with a large number arguments accept description objects encapsulating the arguments instead. 

\begin{lstlisting}[float=t!, language=c++, caption={User-defined types can be used in communication without explicitly creating an MPI data type.}, label=listing:1]
struct custom_type
{
  std::uint64_t        id      ;
  std::array<float, 3> position;
}

custom_type custom;
if (communicator.rank() == 0)
{
  custom = custom_type {42, {1.0f, 2.0f, 3.0f}};
  communicator.send   (custom, 1);
}
if (communicator.rank() == 1)
{
  communicator.receive(custom, 0);
  // custom == custom_type {42, {1.0f, 2.0f, 3.0f}};
}
\end{lstlisting}

User-defined classes must be registered as MPI data types prior to being used in communication.
Our interface is capable of generating MPI data types for custom classes automatically, as seen in Listing \ref{listing:1}.
This functionality is based on PFR \cite{Polukhin2016}, which enables compile-time introspection of aggregate classes.
Arithmetic types, enumerations and specializations of \textit{std::complex} fulfill the \textit{mpi::compliant} concept and are mapped to their MPI equivalents explicitly.
Furthermore, C-style arrays, \textit{std::arrays}, \textit{std::pairs}, \textit{std::tuples} and aggregate types consisting of compliant types are also compliant types themselves.
The communication functions can be used with a single or a contiguous sequential container (i.e. \textit{std::string}, \textit{std::span}, \textit{std::valarray}, \textit{std::vector}) of compliant types.

\begin{lstlisting}[float=b!, language=c++, caption={The requests returned from non-blocking calls can be cast into futures, which can be chained using .then() to express asynchronous sequential operations.}, label=listing:2]
std::int32_t data = 0;
if (communicator.rank() == 0)
  data = 1;

auto status = 
  mpi::future(communicator.immediate_broadcast(data, 0))
  .then([&] (mpi::future f)
  {
    auto status = f.get();
    if (communicator.rank() == 1)
      data++;
    return communicator.immediate_broadcast(data, 1);
  })
  .then([&] (mpi::future f)
  {
    auto status = f.get();
    if (communicator.rank() == 2)
      data++;
    return communicator.immediate_broadcast(data, 2);
  })
  .get(); // data == 3 in all ranks.
\end{lstlisting}

The requests returned by the interface are castable into futures, which can be chained to express asynchronous sequential operations as seen in Listing \ref{listing:2}.
This feature serves as a bridge between the concurrency support library of the C++ standard and the non-blocking communication functionality of MPI.
It further enables task graphs where forks are expressed as multiple futures started from the current context, and joins are expressed with \textit{mpi::when\_all} or \textit{mpi::when\_any} which forward the underlying requests to \textit{MPI\_WaitAll} or \textit{MPI\_WaitAny} respectively.

Error handling is performed by checking the return values of viable MPI functions for success, throwing an exception otherwise.
It is optionally enabled at compile-time by defining a macro prior to inclusion of the library headers.
The exceptions provide an error code, which derives from the error class as specified by the standard.
Default error codes are available as variables scoped in the \textit{mpi::error} namespace.

\section{Performance}
\begin{figure}[t]
  \centering
  \resizebox{0.49\textwidth}{!}{\input{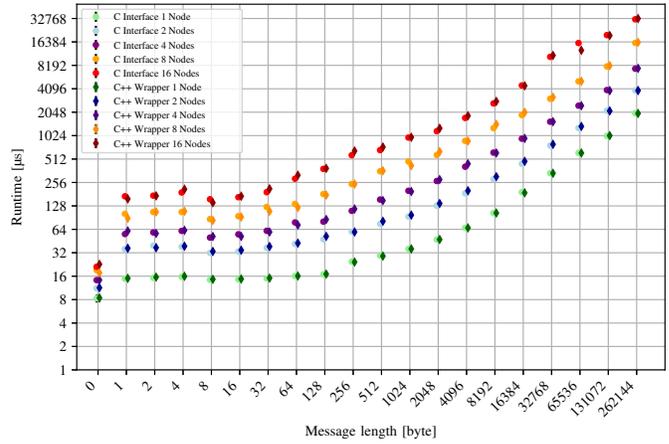}}
  \caption{The benchmark results. The runtime performance of the C and the C++20 interfaces for varying node counts and message lengths.}
  \label{figure:1}
\end{figure}

We measure the performance of the library and compare it to equivalent functionality implemented with the raw C interface.
We utilize mpiBench \cite{mpibench}, which measures the runtime of 11 MPI operations for varying message lengths.
The benchmarks have been adapted to use our interface.

The experiments are controlled by three variables: 
The \textit{interface} varies as C (using the original mpiBench) or C++20 (using the adapted version). 
The \textit{message length} varies as $2^n$ where $0<n<18$. 
The \textit{node count} varies as $1, 2, 4, 8, 16$.
Measurements are taken for each combination of the three variables.
Each measurement is repeated 10 times and averaged.

The experiments are ran on the RWTH Aachen CLAIX-2018 compute cluster.
Each node is equipped with 2 Intel Xeon Platinum 8160 Skylake processors with 24 cores at 2.1 GHz.
The network is provided by a high-speed RDMA Omni-Path interconnect.
The nodes are exclusively reserved for the benchmarks to eliminate effects due to resource consumption of other processes.

The results are shown in Figure \ref{figure:1}.
Each data point represents the geometric mean over the 11 MPI operations.
The slight variances in runtime could be attributed to network traffic which applies even in exclusive mode.
The results of the two implementations do not show recognizable patterns that indicate a disparity in performance.

\section{Conclusion}
We have presented a modern C++ interface for MPI and demonstrated that its performance overhead is negligible in comparison to the raw C interface.
We continue to incorporate the additions and changes that are proposed as part of the 4.1 and 5.0 specifications as they are becoming available.
For further detail, we refer the reader to the source code, distributed under the BSD 3-Clause license, accessible at \url{https://github.com/vrgrouprwth/mpi}.


\bibliographystyle{IEEEtran}
\bibliography{bibliography.bib}
\end{document}